\documentclass{article}
\usepackage{sao2}
\usepackage{psfig}
\newcommand{\be}{\begin{eqnarray}}
\newcommand{\ee}{\end{eqnarray}}
\issue{2005, 58, 118-129}
\begin{document}

\title{Current status of the CATS database}

\author{
O.V.\,Verkhodanov$^{a}$,
S.A.\,Trushkin$^{a}$,
H.\,Andernach$^{b}$,
V.N.\,Chernenkov$^{a}$
}

\institute{
Special Astrophysical Observatory, Nizhnij Arkhyz, Karachaj-Cherkesia,
	 Russia
\and
   Departamento de Astronom{\'{\i}}a,
	Universidad de Guanajuato, Mexico
}

\date{November 3, 2004}{}
\maketitle

\begin{abstract}
We describe the current status of CATS, a publicly accessible database 
(web-server http://cats.sao.ru) allowing one to search in hundreds
of catalogs of astronomical objects discovered all along
the electromagnetic spectrum.
Our emphasis is mainly laid on catalogs of radio continuum sources
observed from 30 to
15000 MHz, secondly on catalogs of objects such as radio
and active stars, planetary
nebulae, HII regions, supernova remnants (SNR), pulsars,
nearby galaxies, AGN and quasars.
CATS also includes the catalogs from the largest extragalactic surveys,
like NVSS, FIRST, WENSS, VLSS, TXS, GB6, SUMSS, IRAS, 2MASS, SDSS,
ROSAT, PGC, MCG, etc.
In 2004 CATS comprised a total of $\sim10^9$ records from over 400
catalogs in the
radio, IR, optical and X-ray windows, including most of RATAN--600 catalogs.
CATS is being expanded and updated, both with newly
published catalogs as well
as older ones which we have created in electronic form for the first time.
We describe the principles of organization of the database of
astrophysical catalogs and the main functions of CATS.

\keywords{astronomical data bases: catalogs ---
radio astronomy: radio sources}

\end{abstract}

\section*{Introduction}

Vast amounts of astrophysical information are now being published, based on
observations of small and large sky regions. Typically, this information
includes coordinates of the observed objects and their physical
characteristics in different wavelength ranges in the form of
source catalogs. In fact, every new large--scale observational experiment
produces a new catalog of objects. Modern
astrophysics operates with source parameters obtained
in different spectral wavelength ranges with the goal of obtaining the most
detailed understanding of physical properties and the processes
of radiation of these objects. The ability of using different
catalogs makes this problem considerably simpler.

  Over the past decades several different attempts have been made
to combine large numbers of astronomical catalogs and make them accessible
in a unified way, which can be classified grossly into two categories:
databases of objects and catalog browsers.
Examples of the former are NED (Helou et al. 1990;
Mazzarella et al. 2002),
Simbad (Egret 1983; Wenger et al. 2000) and LEDA (Paturel et al. 1997).
Examples of the latter are Vizier (Ochsenbein et al. 2000) and AstroBrowser
(http://heasarc.gsfc.nasa.gov/ab/).

  Motivated by
RATAN-600 observation requirements, problems of radio source study, and
the underrepresentation of radio source catalogs in the
then existing object databases, the present authors decided, in 1995,
to create CATS, the {\it Astrophysical {\bf CAT}alogues
{\bf S}upport System}.
Following
the radio astronomical needs, the architecture of operating system and
the considerations expressed in the reviews
by Andernach (1990, 1994, 1999),
we chose to design CATS as a catalog browser rather than an object database,
given that it would allow us to achieve a much better completeness in 
number of records, implying e.g.\ a more complete coverage of radio
source measurements over the entire radio frequency window.
We thus deliberately delayed the known problem of cross-identification
of all the catalogues, as provided in NED, SIMBAD, or LEDA, to a later
stage. In that sense CATS is ideally suited for the experienced researcher
who is looking for the largest amount of data available,
but willing and able
to work out the correct cross-identifications by himself. His effort
will be compensated for by a better completeness of the data than
that obtained from other existing object databases.

  The first steps of the creation of CATS were described by
Verkhodanov \& Trushkin (1994, 1995a,b), Verkhodanov et al. (1997, 2000a)
and Trushkin et al. (2000).
CATS allows one to operate with catalogs stored in ASCII
and FITS Binary tables,
to fit radio continuum spectra and to calculate spectral indices
obtained from
different radio catalogs.
  This database is located at the server {\it cats.sao.ru}
under OS Linux Fedora Core 2, Dual Opteron 244, at Special Astrophysical
Observatory, Russia.

\section*{Implementation of the database}

  The present database consists of catalogs,
their descriptions and corresponding programs operating
under OS Unix.
The program codes are created in the ``C'' algorithmic language
and translated with the GNU project C compiler. The codes
are freely shared, provided they are used for non--commercial goals.
The scheme of the CATS database is shown in Fig.\,1.
\begin{figure*}
\centerline{
\psfig{figure=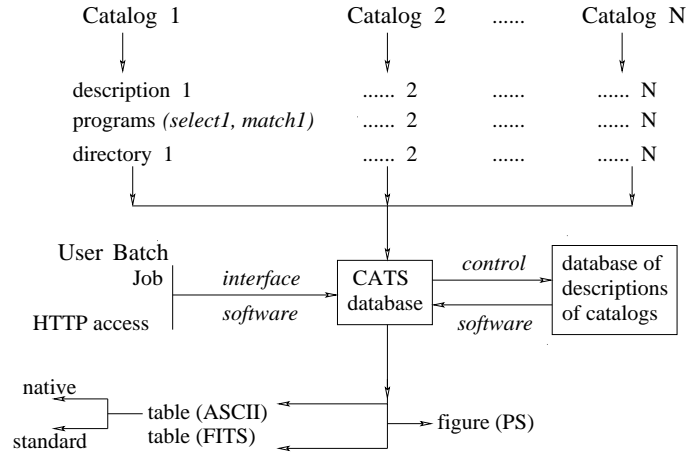,width=9cm,angle=0}
}
\caption{This CATS scheme demonstrates the basic elements of the
database structure: programs and descriptions unified in the corresponding
directory. Information about the directories and programs is stored in
the database of descriptions {\it cats\_descr}. The database
has exchange of information with the repository {\it cats\_descr}.
User requests of selection and matching can be processed in batch mode
or via HTTP access. Output tables are prepared in the ``native''
or ``standard'' format (see text).
}
\end{figure*}

New catalogs may be added to the system in conformity with
the following rules:

1) every new catalog of objects should be placed in
      the UNIX directory with the same name as the catalog of objects;

2) the description of the catalog should also be placed in that
      directory;

3) the programs (or operating scripts)
      for local operations with the catalog
      of objects are also placed in the same directory;

4) brief characteristics, program names and
      description file of the astrophysical catalog are stored
      in the database of descriptions of catalogs named {\it cats\_descr}.

       The following information is stored in the database of
the catalog descriptions {\it cats\_descr\,}: the name of
the catalog, which is coincident with the name of the UNIX
directory, the type of the catalog (radio, optical, combined, etc.),
frequency/wavelength range, minimum fluxes or magnitudes, equatorial
or/and galactic coordinates, names of the local programs
for the ``{\it select}'' and ``{\it match}'' functions
(see below corresponding Section),
the name of the
document file, the number of records in the catalog,
the size of the beam pattern or angular resolution, a recalibration factor
(if available) to put the flux densities on a commonly agreed flux scale,
and a reference.
Parameters from the description file are used by the
programs that process the input data.  E.g., user-specified limits in
coordinates and/or flux density inform the operating programs
about the relevant sky zones to be searched, and thus economize on processing
time if a given catalog is out of the required range.
Whenever a certain catalog column is empty for a given object,
a symbol `n' is returned to the output for the corresponding column,
and considered as an `empty' or `unknown' value.
This symbol is acceptable for the program operating with
this database.

The lower level of the CATS control system includes several basic utilities:
\begin{itemize}
\item
       {\it c\_sel} selects objects with parameters from the given ranges;
\item
       {\it c\_match} looks for objects falling within a certain
  	    coordinate error box;
\item
    {\it cats\_divide} operates with sorted catalogs by right ascension
  	    and produces an index of record numbers corresponding to a certain
	    right ascension;
\item
       {\it epoch} calculates equinox and epoch.
       The epoch is calculated when a given one is different from 1-Jan-2000 or
	1-Jan-1950\footnote{Proper motions are assumed to be zero},
       else the equinox is calculated;
       The catalog is stored in initial equinox haw it was prepared,
	and object coordinates are calculated in to that equinox;
\item
       {\it cats\_base} controls the database of descriptions
  	    {\it cats\_descr} and produces input parameters for the programs
	    {\it c\_sel} and {\it c\_match}.
\end{itemize}
For catalogs in FITS Binary table format (e.g., NVSS or SDSS catalogs),
special programs for the `select' and `match' functions were prepared.

To organize access of a local user of CATS from any directory
of the OS, the operation programs that process the main CATS tasks
are placed in a commonly accessible
directory of the OS UNIX. The control procedures
{\it cats\_sel} and {\it cats\_match}, organized  as shell--scripts,
cover all the low-level programs and provide the interface between CATS and a
local user, or requests made via HTTP or e-mail.
The described method of the catalog storage facilitates
the database development, its expansion with new data
and the possibility to tune the supporting programs.

CATS has its own indexing procedure {{\it cats\_divide}}
for object coordinates in a catalog.
The program decides on where to start searching records of the input
catalog from an index prepared from a right-ascension sorted list.
It allows one to make efficient use of hard-disk seeking functions.
To avoid the problems of searching within a few degrees at the poles
when recalculation from one equinox to another one is done, we process
total list of objects in these zones to find a required source.
The indexing of CATS lists by declination is now under consideration.

\section*{The main functions}

    The following functions are currently implemented in CATS:
\begin{enumerate}
\item Selection of objects from one
      or several catalogs by the main parameters: equatorial or galactic
      coordinates, flux densities, spectral indices, frequencies,
        names, and (for some catalogs)  the type of objects.
        Parameters common to all catalogs (coordinates and flux density)
	are used in selection.
\item Search for counterparts of objects (selected from one or several
	catalogs)
  	by coordinate matching within the error box, circle, or ellipse.
\item Cross--identification of different catalogs.
        This procedure is currently available only for local users.
        It uses the output from the selected zones of one catalog 
        as input for the matching procedure on another catalog.
\item Preparation of receipt with a short description and characteristics
      of each catalog, printing of the total list of catalogs
      for the required sky areas (for local user).
\item Drawing radio spectra of selected sources.
      This can be done from multi-frequency catalogs, or
      catalogs prepared by the CATS team from cross-identification
      of radio catalogs
      at different frequencies, or even by individual user input of
      frequencies and
      flux densities for a source
      (at cats.sao.ru/$\sim$satr/SP/spectrum.html).
      This function is realized in two procedures.
      The first one is a Java-script
      for homogenized CATS data format for single objects
      accessible from the web page. The second one
      is in the
      local data processing system FADPS for object lists.
\end{enumerate}

    The result of the object selection is a datafile
sorted according to different object characteristics
(right ascension (default mode), declination or frequency).
This file can either be displayed on the standard output or sent
to the user in the following formats:
\begin{enumerate}
\item The ``native'' (original) format of the catalog
        (i.e.\ columns as published); for guidance of the user
        a header with a very brief column description is added
        for each catalog that yielded output records.
\item Standard (``homogenized'') output format.
    This format is organized to be
      common for all the catalogs and used  later for unification of
      data, preparation of spectra and statistical analysis.
      The standard FITS Table format describing data and
      fields of the table may be added as a header of
      the resulting file.
\end{enumerate}

  The result may then be used
for subsequent investigation of the radio source spectra or
statistical properties of objetcs in a flexible astronomical
data proccessing system FADPS (Verkhodanov 1997) or, e.g.\ for
the selection of
distant objects in combination with redshift and age estimation
with the system ``Evolution of radio galaxies'' (Verkhodanov et al. 2000b).

\section*{Access to CATS}

Different modes of access are provided according to user
requirements and CATS goals, following modern trends
of software development.
Thus, three main modes of on-line access to the CATS database have
been prepared:
\begin{enumerate}
\item  {\bf Dialogue mode} (non-graphics) has been maintained until
	the present. Several scripts written in the UNIX {\it shell}
       language have been created (Verkhodanov \& Trushkin 1994).
       They permit one to operate with
       the database supporting programs via TCP/IP and NFS
       protocols in the local computer network.
\item  {\bf Hypertext access} (http://cats.sao.ru) is provided
       to allow a user
       from the Internet system to operate with the database CATS
         via hypertext transfer protocol. It allows one
       to execute all described operations from the Web-page.
\item
       {\bf ftp access} (ftp://cats.sao.ru) allows a user to obtain
       both the description of CATS catalogs and the catalogs themselves.
       Here is an example of access by anonymous FTP to the
       catalog WISH:\\
  	{\it anonymous@ftp://cats.sao.ru} \\
	\begin{verbatim}
	ftp>  bin
	ftp>  cd  WISH
	ftp>  get wish11.cat
	ftp>  bye
	\end{verbatim}
  	All the catalog names are stored in the file of descriptions
	{\it cats\_descr} (ftp://cats.sao.ru/cats\_descr).
\item
       {\bf e-mail access} allows the user to send batch
  	requests to CATS.  
	One can send an e-mail with requests of matching with lists or
	selection by several parameters.
	The e-mail  will be read automatically
	and sent for execution to the CATS scripts. The result will be
	sent back automatically to the user via e-mail.
\end{enumerate}

\subsection*{`Select' and `match' procedures}

Two main procedures of data selection have been provided
at different levels of the CATS control.
They are the {\it select} and {\it match}
tasks. They follow the first three main functions described above
and are realized in different access modes.
As was described earlier, the two low-level procedures {\it c\_sel}
and {\it c\_match} provide information according to the corresponding user
requests and pass it to the upper-level control scripts {\it cats\_sel}
and {\it cats\_match}.

  Using these two procedures the cross-identification of catalogs is
possible by using the output from the `select'--function of the
first complete catalog as the input of the `match'--function of the
second catalog or catalogs.

\subsection*{E-mail access}

  In order to economize on the user's real time and avoid the delays
 in on-line data exchange,
we provide a possibility to submit a batch task in the form of an
e-mail letter.

The e-mail requests may have several formats (as explained in a file
that is returned on an empty e-mail to cats@sao.ru, with no subject).
  Two examples of the body of an e-mail
  (no subject required) are shown below.
\begin{itemize}
\item  selection task: \\
    {\it mail -s "" cats@sao.ru }
    \begin{verbatim}
    cats select
    ra min=10:30 max=10:40:00.
    dec > 10' < 12' 30"
    flux > 0.5mJy
    catalogs r equinox=1950
    cats end
    \end{verbatim}
\item  matching task:\\
     {\it mail -s "" cats@sao.ru}
    \begin{verbatim}
    cats match
    catalogs NVSS,FIRST
    window box x=30" y=10'
    sources:
    s1 02:02:00    +31:23:16   1950
    s2 02:23:10     34:03:00   1950
    s3 21:26:33.9  -18:34:33.0 1950
    cats end
    \end{verbatim}
\end{itemize}

These examples demonstrate the use of some keywords for batch requests
(see the results of these requests in Tables 1 and 2).
The beginning
and the end of the request are defined
by the statements: {\tt cats start} and
{\tt cats end}. The first example shows how to search within certain
limits of coordinates by using the ({\tt min, max}) or ($<$, $>$) operators
for right ascension ({\tt ra}) and declination
({\tt dec}).
The keywords expression `{\tt catalogs r}'
sets the type of catalogs to be used for selection, where `{\tt r}' means
all the radio catalogs. Instead of `{\tt r}', one may choose `{\tt o}' for
the optical catalogs, `{\tt x}' for the X--ray ones, `{\tt ir}'
 for the infrared ones, or just the names of used lists separated with comma
(e.g. `{\tt catalogs NVSS,FIRST,WISH}')\footnote{The full list of the
CATS catalogs can be obtained from the CATS Web--page
{\tt http://cats.sao.ru/doc/CATS\_list.html} }.
The keyword {\tt epoch} sets
the equinox of the input coordinates. `{\tt Flux}' sets flux limits.

The second example of the e-mail task is the matching procedure.
There are some additional keywords `{\tt window box x=$30''$ y=$10'$} ',
where `V{\tt box}' is the type of matching window (others are
`{\tt circle}' and `{\tt ellipse}'), `{\tt x}' and `{\tt y}' are
horizontal and vertical coordinate directions, respectively.
Note that {\tt x} and {\tt y} are half
the side lengths of the search {\tt box},
the value following the {\tt circle} keyword is its radius,
and those following
the {\tt ellipse} keyword are the semi-major and semi-minor axes  of
the ellipse.
For the moment~no

\begin{table*}
\caption{Example of an output returned upon a `select' request.
Columns are, respectively, the catalog name, the source name,
the right ascension (hours, minutes and seconds of time) of the object,
the error of RA (in time seconds) if available, otherwise `n',
the declination (degrees, minutes and seconds of arc),
the error of declination (in arcseconds) if available, otherwise `n',
the frequency (MHz), the flux density (Jy), the error of flux density (Jy),
the equinox (`J' means `J2000.0', `B' means `B1950.0').
Note that the names are based on J2000 coords. here, but the user-specified
output
equinox was B1950.0, causing RA and DEC to be different from those
in the J2000-based names.
}
{\small
\begin{verbatim}
# TASK: select
#    default input  epoch: B1950.0
#    default output epoch: B1950.0
#    RA limits:   10:30:00.000 10:40:00.120
#    Dec limits:  00:10:00.001 00:12:29.999
#    GLon limits: 0 360
#    GLat limits: -90 90
#    Flux limits: 0.0005Jy 1000000Jy
#-------------------------------------------
# cat     name            RA          eRA      Dec        eDec freq   Flux(Jy)   eFl equi.
#-----------------------------------------------------------------------------------------
FIRST J103242.4-000331 10 30 08.668      n  +00 11 56.29     n 1400   0.01933 1.38e-04 B
NVSS  J103242-000332   10 30 08.7    0.054  +00 11 55.16  0.89 1400    0.0207    .0005 B
NVSS  J103343-000415   10 31 09.468  0.358  +00 11 14.17 12.24 1400    0.0040    .0006 B
NVSS  J103352-000433   10 31 18.72   0.281  +00 10 56.91  4.23 1400    0.0034    .0005 B
FIRST J103413.2-000432 10 31 39.430      n  +00 10 58.37     n 1400    0.0215 1.47e-04 B
FIRST J103413.2-000442 10 31 39.465      n  +00 10 48.12     n 1400   0.00735 1.46e-04 B
FIRST J103413.2-000453 10 31 39.502      n  +00 10 36.85     n 1400 9.9000e-04 1.46e-04 B
NVSS  J103413-000436   10 31 39.507  0.042  +00 10 54.34  0.73 1400    0.0317    .0005 B
FIRST J103748.6-000515 10 35 14.872      n  +00 10 20.65     n 1400   0.01031 1.48e-04 B
FIRST J103749.3-000522 10 35 15.577      n  +00 10 14.02     n 1400    0.0038 1.48e-04 B
NVSS  J103749-000521   10 35 15.601  0.052  +00 10 14.49  0.81 1400    0.0320    .0014 B
FIRST J103750.1-000525 10 35 16.372      n  +00 10 10.34     n 1400   0.01009 1.48e-04 B
FIRST J103846.6-000316 10 36 12.838      n  +00 12 21.62     n 1400   0.00132 1.38e-04 B
NVSS  J103749-000521   10 35 15.601  0.052  +00 10 14.49  0.81 1400    0.0320    .0014 B
FIRST J103750.1-000525 10 35 16.372      n  +00 10 10.34     n 1400   0.01009 1.48e-04 B
FIRST J103846.6-000316 10 36 12.838      n  +00 12 21.62     n 1400   0.00132 1.38e-04 B
NVSS  J103950-000324   10 37 16.805  0.598  +00 12 14.43  5.87 1400    0.0038    .0011 B
NVSS  J104008-000354   10 37 35.191  0.077  +00 11 45.81  1.27 1400    0.0134    .0005 B
FIRST J104008.9-000353 10 37 35.219      n  +00 11 45.97     n 1400   0.01254 1.47e-04 B
FIRST J104220.5-000409 10 39 46.783      n  +00 11 33.46     n 1400   0.00103 1.41e-04 B
#----------------------------------------------------------------
#The catalogue identifications listed are related to the following references:
#----------------------------------------------------------------
# FIRST  : 1997ApJ...475..479White+ FIRST survey catalogue at 1.4GHz
# NVSS   : 1998AJ....115.1693Condon+ 1996: NVSS survey catalog
\end{verbatim}
}
\end{table*}

\begin{table*}
\caption{Example of output data returned upon a `match' request.
The first 14 columns correspond to the columns of the Table 1.
Other columns are, respectively,
the distance from the search position (in arcsec),
the position angle (in degrees
from N through E) of the radius vector connecting the search center with the
object.
}
{\small
\begin{verbatim}
# TASK: match
# Search box 30 x 600 arcsec
# default input  epoch: J2000.0
# default output epoch: J2000.0
#
# cat     name            RA         eRA       Dec       eDec  freq Flux(Jy)  eFl equi.dist," pa,^
#-------------------------------------------------------------------------------------------------
#OBJECT: s1 02:02:00 +31:23:16 1950
WENSS WNB0202.5+3124   02 05 29.585      n  +31 39 07.7      n  325   0.077  .0036 J   453.9  282
NVSS  J020529+313912   02 05 29.738  0.071  +31 39 12.15  0.98 1400  0.0170  .0005 J   456.7  282
#@-------------------------------------------
#OBJECT: s2 02:23:10 34:03:00 1950
NVSS  J022526+341006   02 25 26.717  0.374  +34 10 06.63  3.85 1400  0.0037  .0005 J   664.4  125
NVSS  J022529+342450   02 25 29.186  0.412  +34 24 50.47   5.6 1400  0.0030  .0007 J   716.3   46
WENSS WNB0223.1+3408   02 26 10.154      n  +34 21 30.7      n  325   3.678  .0045 J   301.8    1
GB6   J0226+3421       02 26 10.2      0.5  +34 21 25        8 4850   1.628   .145 J   296.1    1
NVSS  J022610+342130   02 26 10.337  0.036  +34 21 30.31  0.56 1400  2.8949  .0005 J   301.4    0
NVSS  J022649+340740   02 26 49.391  0.349  +34 07 40.85   4.2 1400  0.0034  .0007 J   715.4  222
#@-------------------------------------------
#OBJECT: s3 21:26:33.9 -18:34:33.0 1950
NVSS  J212840-183008   21 28 40.07   0.272  -18 30 08.45  6.54 1400  0.0039  .0006 J   788.5  228
NVSS  J212840-181420   21 28 40.26   0.306  -18 14 20.93  3.08 1400  0.0048  .0005 J   722.6  306
NVSS  J212840-182926   21 28 40.26   0.537  -18 29 26.22  6.54 1400  0.0034  .0006 J     759  230
NVSS  J212841-181322   21 28 41.531  0.299  -18 13 22.25  5.09 1400  0.0049  .0006 J   744.6  310
NVSS  J212855-181737   21 28 55.425  0.035  -18 17 37.7   0.78 1400  0.1092  .0040 J   433.5  301
NVSS  J212859-181253   21 28 59.241  0.038  -18 12 53.21  0.65 1400  0.0462  .0018 J     600  328
NVSS  J212921-182122   21 29 21.414  0.032  -18 21 22.99  0.56 1400  1.4124  .0005 J     0.3  337
NVSS  J212945-182052   21 29 45.696  0.112  -18 20 52.35  1.61 1400  0.0089  .0005 J   346.9   85
NVSS  J212950-181500   21 29 50.726  0.387  -18 15 00.03  5.15 1400  0.0032  .0005 J   566.6   47
NVSS  J212952-182910   21 29 52.722  0.045  -18 29 10.27   0.7 1400  0.0372  .0015 J   645.2  136
#@-------------------------------------------
#The catalogue identifications listed are related to the following references:
#----------------------------------------------------------------
# GB6    : 1996ApJS..103..427Gregory+ The GB6 catalog;;
# NVSS   : 1998AJ....115.1693Condon+ 1996: NVSS survey catalog (updated! - v.40, Jul-02);
# WENSS  : 1997A&AS..124..259Rengelink+ The Westerbork Northern Sky Survey (WENSS).
\end{verbatim}
}
\end{table*}

\clearpage
\noindent  position angle (PA) of the search ellipse
may be given, and PA = 0$^{\circ}$ is assumed (i.e.\ semi-major axis along
the N--S direction).
The keyword `{\tt sources:}' shows that the following records separated by
a newline code are source coordinates to be used for the matching procedure.
Each record contains the name of an object, its equatorial coordinates (R.A.
and Dec.) and the corresponding equinox.

Keywords in the batch request may be separated by a space, a tabulation
or newline character.

The full information about the keywords and formats can be requested
with an empty e-mail to the address {\tt cats@sao.ru} (no subject required).

  Results of these two sample requests are shown in Tables 1 and 2.

\section*{The main catalogs}
  The major source of radio catalogs of CATS is the collection of 
one of us (Andernach 1990, 1999), who has spent
large efforts to recover older source lists not previously available
in electronic form, using a scanner and optical character recognition
software.
This collection is complemented
in several ways: contributions from authors, astro-ph preprints,
tables from electronic journals and the CDS catalog archive, as well as
occasional manual retyping of the original catalog.
About 70 catalogs have been typed and/or corrected manually
at SAO RAS by S.Trushkin.
The largest catalogs
(e.g. NVSS, FIRST, SDSS, etc.) were copied from Web-sites of the catalog
authors.

  CATS is mainly a radio-astronomical database. All major catalogs
incorporated into CATS are shown in Tables 3 and 4 (adapted and updated from
Table~1 of Andernach 1999).

{\small
\begin{onecolumn}
\makebox[15cm]{}
\begin{center}
Table 3:{ \it
Major radio astronomical catalogs of the CATS database}
\medskip
\tablehead{
\hline
	\multicolumn{1}{|c|}{Freq.}
      & \multicolumn{1}{|c|}{ }
      & \multicolumn{1}{|c|}{Year}
      & \multicolumn{1}{|c|}{RA (h)}
      & \multicolumn{1}{|c|}{Dec (\degr)}
      & \multicolumn{1}{|c|}{HalfPower}
      & \multicolumn{1}{|c|}{S$_{min}$}
      & \multicolumn{1}{|c|}{Number}
      \\
	\multicolumn{1}{|c|}{ }
      & \multicolumn{1}{|c|}{Name}
      & \multicolumn{1}{|c|}{of}
      & \multicolumn{1}{|c|}{ }
      & \multicolumn{1}{|c|}{ }
      & \multicolumn{1}{|c|}{BeamWidth}
      & \multicolumn{1}{|c|}{ }
      & \multicolumn{1}{|c|}{of}
      \\
	\multicolumn{1}{|c|}{(MHz)}
      & \multicolumn{1}{|c|}{ }
      & \multicolumn{1}{|c|}{publ.}
      & \multicolumn{1}{|c|}{or l (d)}
      & \multicolumn{1}{|c|}{or b (d)}
      & \multicolumn{1}{|c|}{(arcmin)}
      & \multicolumn{1}{|c|}{(mJy)}
      & \multicolumn{1}{|c|}{objects}
      \\ \hline
}
\tabletail{
\hline}
\par
\begin{supertabular}{|rcrcccrr|}
 10-25& UTR-2  & 78-95 &   0-24   &   $>$-13    & 25..60  & 10000 &   1754 \\
   38 & 8C     & 90/95 &   0-24   &   $>$+60    &   4.5   & 1000  &   5859 \\
   74 & VLSS   & 2004  &   0-24   &   $>-30$    &   1.3   &  350  &  32521 \\
   80 & CUL1   &  73   &   0-24   &    -48,+35  &   3.7   & 2000  &    999 \\
   80 & CUL2   &  75   &   0-24   &    -48,+35  &   3.7   & 2000  &   1748 \\
   82 & IPS    &  87   &   0-24   &    -10,+83  & 27x350  &  500  &   1789 \\
  151 & 6CI    &  85   &   0-24   &    $<$+80   &   4.5   &  200  &   1761 \\
  151 & 6CII   &  88   &  8.5-17.5&    +30,+51  &   4.5   &  200  &   8278 \\
  151 & 6CIII  &  90   &  5.5-18.3&    +48,+68  &   4.5   &  200  &   8749 \\
  151 & 6CIV   &  91   &   0-24   &    +67,+82  &   4.5   &  200  &   5421 \\
  151 & 6CVa   &  93   &  1.6- 6.2&    +48,+68  &   4.5   & ~300  &   2229 \\
  151 & 6CVb   &  93   & 17.3-20.4&    +48,+68  &   4.5   & ~300  &   1229 \\
  151 & 6CVI   &  93   & 22.6- 9.1&    +30,+51  &   4.5   & ~300  &   6752 \\
  151 & 7CI    &  90   & (10.5+41)&    (6.5+45) &   1.2   &   80  &   4723 \\
  151 & 7CII   &  95   &   15-19  &    +54,+76  &   1.2   & ~100  &   2702 \\
  151 & 7CIII  &  96   &   9-16   &    +20,+35  &   1.2   & ~150  &   5526 \\
  160 & CUL3   &  77   &   0-24   &    -48,+35  &   1.9   & 1200  &   2045 \\
  178 & 4C     &  65   &   0-24   &    -7,+80   & 15x7.5  & 2000  &   4844 \\
  232 & MIYUN  &  96   &   0-24   &    +30,+90  &   3.8   & ~100  &  34426 \\
  325 & WENSS  &  98   &   0-24   &    +30,+90  &   0.9   &   18  & 229420 \\
  327 & WSRT   &  91   &  5 fields&   (+40,+72) &  ~1.0   &    3  &   4157 \\
  352 & WISH   & 2002  &   0-24   &    -25,-9   &   0.9   &   18  &  90357 \\
  365 & TXS    &  96   &   0-24   &  -35.5,+71.5&   ~.1   &  250  &  66841 \\
  408 & MRC    & 81/91 &   0-24   &   -85,+18.5 &   ~3    &  700  &  12141 \\
  408 & B2     & 70-73 &   0-24   &    +24,+40  &  3 x10  &  250  &   9929 \\
  408 & B3     &  85   &   0-24   &    +37,+47  &  3 x 5  &  100  &  13354 \\
  608 & WSRT   &  91   & sev.fields&  (~40,~72) &   0.5   &    3  &  1693  \\
  611 & NAIC   &  75   &   22-13  &    -3,+19   &  12.6   &  350  &   3122 \\
  843 & SUMSS  &  99   &   0-24   &    $<-30$   &   0.72  &    6  & 178975 \\
 1400 & GB     &  72   &   7-16   &    +46,+52  &  10x11  &   90  &   1086 \\
 1400 & GB2    &  78   &   7-17   &    +32,+40  &  10x11  &   90  &   2022 \\
 1400 & WB92   &  92   &   0-24   &    -5,+82   &  10x11  & ~150  &  31524 \\
 1400 & NVSS   &  98   &   0-24   &    -40,+90  &   0.9   &  2.0  &1810668 \\
 1400 & FIRST  &  98   &  7.3,17.4&   22.2,57.6 &   0.1   &  1.0  & 811117 \\
      &        &  98   & 21.3,3.3 &   -11.5,+1.6&         &       &        \\
 1400 & PDF    &  98   & 1.1 -1.3 &    -46,-45  & 0.1-0.2 &  0.1  &   1079 \\
 1500 & VLANEP &  94   & 17.4,18.5&   63.6,70.4 &  0.25   &  0.5  &   2436 \\
 2700 & PKS    & (90)  &   0-24   &    -90,+27  &   ~8    &   50  &   8264 \\
 3900 & Z      &  89   &   0-24   &     0,+14   & 1.2x52  &   50  &   8503 \\
 3900 & RC     & 91/93 &   0-24   &    4.5,5.5  & 1.2x52  &    4  &   1189 \\
 3900 & Z2     &  95   &   0-24   &     0,+14   & 1.2x52  &   40  &   2943 \\
 4850 & MG1-4  & 86-91 &   var.   &     0,+39   &  ~3.5   &   50  &  24180 \\
 4850 & 87GB   &  91   &   0-24   &     0,+75   &  ~3.5   &   25  &  54579 \\
 4850 & GB6    &  96   &   0-24   &     0,+75   &  ~3.5   &   18  &  75162 \\
 4850 & PMNM   &  94   &   0-24   &    -88,-37  &   4.9   &   25  &  15045 \\
 4850 & PMN-S  &  94   &   0-24   &   -87.5,-37 &   4.2   &   20  &  23277 \\
 4850 & PMN-T  &  94   &   0-24   &    -29,-9.5 &   4.2   &   42  &  13363 \\
 4850 & PMN-E  &  95   &   0-24   &    -9.5,+10 &   4.2   &   42  &  11774 \\
 4850 & PMN-Z  &  96   &   0-24   &    -37,-29  &   4.2   &   70  &   2400 \\
\hline
\hline
   31 & NEK    &  88   &$350<l<250$& $ |b|<2.5$ &  13x 11 & 4000  &    703 \\
  151 & 7C(G)  &  98   &$80<l<180 $& $ |b|<5.5$ &   1.2   & ~100  &   6262 \\
  327 & WSRTGP &  96   &$ 43<l<91 $& $ |b|<1.6$ &  ~1.0   &  ~10  &   3984 \\
 1400 & GPSR   &  90   &$ 20<l<120$& $ |b|<0.8$ &  0.08   &   25  &   1992 \\
 1408 & RRF    &  90   &$357<l<95.5$&$ |b|<4.0$ &   9.4   &   98  &    884 \\
 1420 & RRF    &  98   &$95.5<l<240$&$ -4<b<5 $ &   9.4   &   80  &   1830 \\
 1400 & GPSR   &  92   &$350<l<40 $& $ |b|<1.8$ &  0.08   &   25  &   1457 \\
 2700 & F3R    &  90   &$357<l<240$& $ |b|< 5 $ &   4.3   &   40  &   6483 \\
 4875 & ADP79  &  79   &$357<l<60 $& $  |b|<1 $ &   2.6   & ~120  &   1186 \\
 5000 & GT     &  86   &$40<l<220 $& $  |b|<2 $ &   2.8   &   70  &   1274 \\
 5000 & GPSR   &  94   &$350<l<40 $& $ |b|<0.4$ &  ~0.07  &    3  &   1272 \\
 5000 & GPSR   &  79   &$190<l<40 $& $  |b|<2 $ &   4.1   &  260  &    915 \\
\end{supertabular} 
\end{center}
\end{onecolumn}

\setcounter{table}{3}

\begin{table*}[!th]
\begin{center}
\caption{Some catalogs of other wavelength ranges in the CATS}
\medskip
\begin{tabular}{|cccccr|}
\hline
$\lambda$& PGC & Publ &    RA    &      Dec    &  Number   \\
\hline
 opt  & PGC    &  89   &   0-24   &    -90,+90  &   73197 \\
 opt  & MCG    &  75   &   0-24   &    -33.5,+90&   31886 \\
 opt  & MSL    &  85   &   0-24   &    -90,+90  &  181603 \\
 opt & SDSS DR2 & 2004 &  several &             & Gals: 216906 \\
      &        &       &  fields  &            &   QSOs: 22033 \\
 ir   &IRASPSC &  87   &   0-24   &    -90,+90  &  245889 \\
 ir   &IRASFSC &  89   &   0-24   & $|b|>10$    &  235935 \\
 ir   &IRASSSC &  89   &   0-24   &    -90,+90  &   43886 \\
 ir   & 2MASS  & 2000  &  0-24    &   -90,+90   &  470992970 \\
 Xray & ROSAT  &  95   &   0-24   &    -90,+90  &   74301 \\
 mix  &QSO HB2 &  93   &   0-24   &    -84,+86  &    7315 \\
 mix  &VERON+11&  93   &   0-24   &    -83,+85  &   48921 \\
\hline
\end{tabular}
\end{center}
\end{table*}
}
Names of catalogs contained in CATS with corresponding references are
presented in {\bf Appendix}.

CATS also contains observational and combined catalogs of Galactic supernova
remnants (e.g., Trushkin et al. 1987; Trushkin 1996),
secondary tables with objects
selected by spectral index
(Chambers et al. 1996; R\"ottgering et al. 1994; De Breuck et al. 2000),
variability or AGN properties (Kovalev et al. 1999, 2000).
  Also the database includes source catalogs obtained 
  from cross-identifications performed within CATS
(IRAS--TXS: Trushkin \& Verkhodanov (1995) and
Verkhodanov \& Trushkin (2000); UTR\_ID: Verkhodanov et al. (2000c, 2003);
WMAP: Trushkin (2004)).

%
%

\section*{Examples of some typical tasks}

\begin{itemize}
\item {\it Extraction of a sample of steep-spectrum radio sources
   from the FIRST catalog}. \\
   To execute this task, one may use the FIRST (1400 MHz) and Texas
   (365 MHz) catalogs selection in a given zone.
   The FIRST and Texas catalogs have a flux density limit
   of about 1 and 150 mJy, respectively.
   A cross-matching of objects in any sky zone where these two catalogs
   overlap will return sources with the required properties.
   Thus, we prepare 2 requests: (1) select all objects from
   a given zone in the FIRST catalog with the `select'--function,
   (2) using the list of FIRST sources obtained in (1)
   we cross-identify selected objects with
   the Texas data with the `match'--function.
   The resulting list can be used to finish the required selection in the
   given zone.
\item {\it Identification of objects with optical catalogs in CATS}.\\
	Open the CATS web page with optical catalogs, and provide the input
       list of objects. Alternatively,
	select `{\tt catalogs o}'
	in your e-mail request, provide a list of objects of interest,
  	and send the request.
\end{itemize}

\vspace*{-0.1cm}
\section*{Summary}
      CATS provides a simple and convenient access to astrophysical data
and complements the data available from other services, most notably
for radio continuum flux densities, for which it is the largest database
in existence.  Operation with the database permit
  astronomers to search for peculiar objects and study physical
  processes in sources of cosmic radiation.

        Until October 2004, we registered over 28000
  requests for the `select' or `match' procedures, which are
the most popular.
The most popular
catalogs for FTP-copying over the last five years are QSO by Hewitt \&
Burbidge (20 times), PGC (19 times), and NVSS (18 times). CATS processes
daily up to 1000 HTTP--requests for information concerning
the catalog descriptions.

   CATS is being expanded continuously and presently comprises more than
400 catalogs including all the RATAN--600 catalogs.
  The database in its present form occupies about 60\,Gb of disk space.

\vspace*{-0.1cm}
\begin{acknowledgements}
The authors are thankful to Alexander Kopylov and
Grigorij Tsarevsky for testing CATS procedures and catalogs and
suggesting some improvements, and to N.F.\,Voikhanskaya for useful
remarks. This work is supported by the grants
96--07--89075, 02--07--90038 of Russian
Foundation of Basic Researches.
\end{acknowledgements}

\newpage
\begin{table*}
\begin{center}
{\bf \large Appendix}
\begin{tabular}{ll}
&\\
UTR-2  & Braude et al. (1978, 1979, 1981, 1985, 1994, 2002);\\
8C &     Hales et al. (1995);\\
VLSS  &  Lane et al., (2004);\\
CUL1,2,3 & Slee et al. (1995);\\
IPS &    Purvis et al. (1987),\\
6C &     Baldwin et al. (1985), Hales et al. (1988, 1990, 1991, 1993a,b, 1995);\\
7C&      McGilchrist \& Riley (1990), Visser et al. (1995), Lacy et al. (1995),\\
&         Waldram et al. (1996), Pooley et al. (1998), Riley et al. (1999a,b);\\
4C &     Gower et al. (1967);\\
MIYUN&   Zhang et al. (1997);\\
WENSS &  Rengelink et al. (1997);\\
WSRT &   Valentijn et al. (1977), Goss et al. (1977, 1980), Wouterloot \& Dekker (1979),\\
&         Habing et al. (1982), Isaacman (1981), Matthews \&  Spoelstra (1983),\\
&         Oort \& Windhorst (1985), Wieringa (1991), Roland et al. (1990), Taylor et al. (1996); \\
WISH&     De Breuck et al. (2002);\\
TXS&      Douglas et al. (1996);\\
MRC &     Large et al. (1991);\\
B2&       Colla et al. (1970);\\
B3&       Ficarra et al. (1985);\\
NAIC &    Durdin et al. (1975), Lawrence et al. (1986);\\
SUMSS&    Bock et al. (1999);\\
GB &      Maslowski (1972), Machalski (1978), Rys \& Machalski (1987);\\
WB92 &    White \& Becker (1992);\\
NVSS&     Condon et al. (1998);\\
FIRST&    White et al. (1997)\\
PDF&      Hopkins (1998), Hopkins et al. (1998);\\
VLANEP&   Kollgaard et al. (1994);\\
PKS&      Wright \& Otrupcek (1990), Otrupcek \& Wright (1991);\\
Z&        Amirkhanyan et al. (1989);\\
RC&       Parijskij et al. (1991, 1992);\\
Z2&       Larionov et al. (1991);\\
MG1-4&    Bennett et al. (1986), Langston et al. (1990), Griffith et al. (1990, 1991); \\
87GB &    Gregory \& Condon (1991);\\
GB6 &     Gregory et al. (1996);\\
PMN&      Gregory et al. (1994), Wright et al. (1996);\\
NEK&      Kassim (1988);\\
7C(G)&    Vessey \& Green (1998);\\
WSRTGP&   Taylor et al. (1996);\\
GPSR &    Becker et al. (1994), Garwood et al. (1988), Helfand et al. (1992), Zoonematkermani \\
& et al. (1990);\\
RRF &     Reich et al. (1990);\\
F3R &     F\"urst et al. (1990);\\
ADP79&    Altenhoff et al. (1979)\\
GT&       Gregory \& Taylor (1986);\\
 PGC&     Paturel et al. (1989);\\
 MCG &    Kogoshvili (1982);\\
 MSL &    Dixon (1970, 1980), Dixon et al. (1981), with corrections from Andernach (1989) and \\
& later updates;\\
SDSS DR2& Schneider et al. (2001, QSOs list), Abazajian et al. (2004);\\
IRASPSC & Beichman et al. (1988): IRAS Point Source Catalog;\\
IRASFSC&  Beichman et al. (1988): IRAS Faint Source Catalog;\\
IRASSSC & Beichman et al. (1988): IRAS Serendipitous Survey Catalog;\\
 2MASS &  Cutri et al. (2002);\\
 ROSAT &  White et al. (2000);\\
QSO HB &  Hewitt \& Burbidge (1993);\\
VERON+11& V\'eron-Cetty and V\'eron (2003).\\

\end{tabular}
\end{center}
\end{table*}

\end{document}